\documentclass[twocolumn,showpacs,prl]{revtex4-1}

\usepackage{amsmath}
\usepackage{amsfonts}
\usepackage{amssymb}
\usepackage{graphics} 
\usepackage{graphicx}

\newcommand{\eps}{\varepsilon}

\begin{document}
\title{Helical liquid of snake states}

\author{Kh.~Shakouri}
\affiliation{Department of Physics, University of Antwerp, Groenenborgerlaan 171, B-2020 Antwerpen, Belgium}

\author{S.~M.~Badalyan}
\email{Samvel.Badalyan@ua.ac.be}
\affiliation{Department of Physics, University of Antwerp, Groenenborgerlaan 171, B-2020 Antwerpen, Belgium}

\author{F.~M.~Peeters}
\affiliation{Department of Physics, University of Antwerp, Groenenborgerlaan 171, B-2020 Antwerpen, Belgium}

\begin{abstract}
We derive an exact solution to the problem of spin snake states induced in a nonhomogeneous magnetic field by a combined action of the Rashba spin-orbit and Zeeman fields. In an antisymmetric magnetic field the spin snake states are nonlocal composite particles, originating from spatially separated entangled spins. Adding an external homogeneous magnetic field breaks the spin-parity symmetry gapping out the spectral branches, which results in a regular beating pattern of the spin current. These new phenomena in a helical liquid of snake states are proposed for an experimental realization.
\end{abstract}


\date{\today}

\maketitle

In a spatially nonhomogeneous magnetic field the alternating Lorenz force produces opposite rotations of electrons on both sides of the magnetic interface, resulting in a drift of carriers in snake-like classical trajectories. Gapless excitations of snake states are formed, leading to conduction along the magnetic interface. The snake states have been a subject of active theoretical \cite{Muller1992,Peeters1993,Peeters1999,SMB2001} and experimental \cite{Carmona1995,Ye1995,Nogaret2000} investigations in two-dimensional electron systems (2DES) formed in semiconductor heterojunctions as well as in graphene \cite{Oroszl,Ghosh2008,Martino2011}.
Snake states formed in homogeneous magnetic fields along the charge neutrality lines of electrons and holes play a key role in ambipolar transport of topological insulators \cite{Gusev2010,Gusev2012} and graphene \cite{Marcus2011,Sun201212}.

Spin-orbit interaction (SOI) provides an effective tool for an electrical control of the spin behavior \cite{Fabian2007}. It enables physical separation of the spin of charge carriers \cite{Dyakonov1971,SMB2009,Glazov2011}, permitting spin transport properties in a non-ohmic regime. Helical one-dimensional states with opposite spins moving in opposite directions can be effectively generated by a combined action of SOI and a homogeneous magnetic field in quantum wires \cite{Streda2003,Debald2005,Klinovaja2012} and at edges of 2DES \cite{Scharf2012,SMB2010,SMB200909}. Suppression of backscattering processes and a large inter-edge states equilibration length \cite{Komiyama1992,SMB1991} make such spin edge states suitable for spintronics applications \cite{Ji2003,Nakajima2012}, particularly for quantum information processing purposes \cite{Bertoni2000,Chirolli2012}.
As a distinction from spin edge states, the helical system of snake states exhibits unidirectional motion, which emerges due to the spatially nonhomogeneous magnetic field, and which in a combination with SOI can produce additional spin transport channels with even richer physics.
 
Here we present an exact solution of the problem of spin snake states that emerge due to combined action of the Rashba spin-orbit and Zeeman fields in the presence of a magnetic interface. Our theory is equally applicable to the spin snakes formed along p-n junctions in a homogeneous magnetic field. We find that the energy dispersions are described by the main Landau number, the spin-parity and the spin-chirality quantum numbers. In an antisymmetric magnetic field the spin-split quasibulk Landau levels are degenerate with respect to the spin-parity. Close to the magnetic interface this degeneracy is lifted except the point where the even and odd spin-parity terms cross. 
We demonstrate that for such a magnetic field profile the electron spin vanishes for states located far from the magnetic interface. This implies that an electron in such a spin snake state is transformed into a composite particle, which experiences simultaneously positive and negative magnetic fields. Such a {\it nonlocal} composite particle executes semiclassical cyclotron orbits with opposite rotations on both sides of the magnetic interface, producing {\it entangled} spins separated in space. When an additional homogeneous magnetic field is applied, moving the total averaged magnetic field slightly away from zero, the spin-parity symmetry breaks down and we find that the electron spin makes a jump, acquiring a finite out-of-plane component. Towards the magnetic interface the in-plane x-component (perpendicular to the interface) of the electron spin becomes finite while the out-of-plane z-component decreases. Both spin projections show an oscillating behavior before the z-component will tend to zero and the x-component$-$to its quantized value. This is consistent with changes of the Rashba field in parallel to the transformation of the electron circular motion into an unidirectional one along the magnetic interface. Furthermore, at the energies corresponding to the crossings in the spectrum of the spin snake states, we find strong peaks in the spin current. Applying an external uniform magnetic field gaps out the electronic spectral branches and produces a helical liquid of spin snake states resulting in a regular {\it beating} patterns in the spin current. These  predictions of the properties of spin snake states can be used in an experimental probe of this special class of helical liquid and be of practical relevance for quantum information processing and spintronic applications.

Consider a 2DES formed in a quantum well parallel to the (001) plane of a zinc-blende semiconductor heterostructure. A nonhomogeneous antisymmetric magnetic field, $\text{sign}(x) B_{0}$, together with an external uniform field, $B_u$, is applied perpendicular to the 2DES. In the presence of the Rashba spin-orbit and Zeeman fields, the effective Hamiltonian of the system can be written as
\begin{eqnarray}\label{H}
	H=&&\frac{\vec{\pi}^2}{2m^\ast}\openone +\alpha\left[\pi_x\hat\sigma_y-\pi_{y}\hat\sigma_x\right] -\frac{1}{2}g\mu_B B(x) \hat\sigma_z.
\end{eqnarray}
Here $m^\ast$ is the electron effective mass, $\vec{\pi}$ the kinetic momentum, $\openone$ and $\hat\sigma_{x,y,z}$ the unit and Pauli matrices in spin space, $\mu_B$ the Bohr magneton, $\alpha$ and $g$ the strength of Rashba SOI and the Land$\grave{e}$ factor, respectively. In the Landau gauge with the vector potential $\vec{A}=(0, x B(x))$ where $B(x) =B_u+ \text{sign}(x) B_0$ is the total magnetic field, the Hamiltonian (\ref{H}) is invariant under translations along the $y$ direction and we can reduce the two-dimensional Schr{\"{o}}dinger equation into an effective one-dimensional problem making use of the ansatz $\Psi(x,y)=e^{i k_y y}\chi_{k_y}(x)$. Here $k_y$ is the electron wave vector and $ \chi_{k_y}(x)=\left| \chi_{1 k_y}(x), \chi_{2 k_y}(x) \right|^{T}$ the spinor wave function. We introduce dimensionless quantities for the x-coordinate, $x/l_B/\sqrt{2}$, and for the center of cyclotron orbit, $X(k_y)=\sqrt{2}k_yl_B$, in units of the magnetic length, $l_B=\sqrt{\hbar/e B}$, and for the electron energy, $\eps /\hbar\omega_{B}=\nu+1/2$, in units of the cyclotron frequency, $\omega_{B}=e B/m^\ast$. Here $B=B_{u}+B_{0}$ and it is assumed that $B_{u}, B_{0}>0$. Then we can cast the Schr{\"{o}}dinger equation for $\chi_{k_y}(x)$ in the following compact matrix form 
\begin{equation}\label{compct}
\left(
\begin{array}
[c]{cc}
h_{\nu \pm \lambda} & \pm h_{\pm}\\
\pm h_{\mp} & h_{\nu \mp \lambda}%
\end{array}
\right)  \left(
\begin{array}
[c]{c}%
\chi_{1k_{y}}\left(  \xi\right) \\
\chi_{2k_{y}}\left(  \xi\right)
\end{array}
\right) =0
\end{equation}
where  $\xi=x \sqrt{\theta(x)} \mp X(k_{y})/\sqrt{\theta(x)}$ with the upper (lower) sign corresponding to $x>0$ ($x<0$). The step function $\theta(x)=1$ for $x>0$ and $\theta(x)=|B_{u}-B_{0}|/(B_u+B_0)$ for $x<0$. The operators in Eq.~(\ref{compct}) are defined as
\begin{eqnarray}\label{oper}
h_{\nu\pm\lambda}&=& \frac{d^{2}}{d\xi^{2}}+\frac{1}{\theta(x)}\left(\nu+\frac{1}{2}\right) \pm \lambda -\frac{\xi^{2}}{4}~,
\\
h_{\pm}  &=&-\frac{\gamma}{\sqrt{\theta(x)}}\left(  \frac{\xi}{2}\mp\frac{d}{d\xi}\right) \nonumber
\end{eqnarray}
where the dimensionless parameters $\gamma=2\alpha/v_B$ ($v_{B}=\sqrt{2}l_{B}\omega_{B}$ is the cyclotron velocity)  
and $\lambda=g\mu_B m^\ast /2 e\hbar$ describe the spin-orbit and Zeeman fields, respectively, and $\gamma$ depends on the magnetic field strength. 

In the absence of SOI, $h_{\pm}=0$, the solution of (\ref{compct}) is given in terms of the scalar parabolic cylindrical functions, $D_{\nu}(x)$ \cite{SMB2001}. For $\nu\geqslant-1/2$ and $\nu\neq 0,1,2,...$ there exist two independent solutions, $D_{\nu}(x)$ and $D_{\nu}(-x)$ and $D_{\nu}(x) \rightarrow 0$ only for $x \rightarrow+\infty $. Making use of this behavior, we construct the solution of the matrix equation (\ref{compct}) satisfying the boundary conditions at infinity in terms of the following spinors
\begin{equation}\label{bulk}
\chi_{\pm}^{L}(\xi)=\left(
\begin{array}
[c]{c}%
c^{L}_{\pm}D_{\mu^{L}_{\pm}-1}\left(-\xi^{L}\right) \\
D_{\mu^{L}_{\pm}}\left(-\xi^{L}\right)
\end{array}
\right),~
\chi_{\pm}^{R}(\xi)=\left(
\begin{array}
[c]{c}%
D_{\mu^{R}_{\pm}}\left(\xi^{R}\right) \\
c^{R}_{\pm}D_{\mu^{R}_{\pm}-1}\left(\xi^{R}\right)
\end{array}
\right)~,
\end{equation}
which represent the bulk solutions of Eq.~(\ref{compct}) when
\begin{eqnarray}\label{indec}
\mu^{L,R}_{\pm}&=&\frac{1}{\theta(x)}\left(\nu+\frac{1}{2}+\frac{\gamma^{2}}{2} \pm \sqrt{\cal D}\right)~,
\\
c^{L,R}_{\pm}&  =&-\frac{\sqrt{\theta(x)}}{\gamma}\left(\frac{1}{2}-\lambda +\frac{1}{\theta(x)}\left(\frac{\gamma^{2}}{2} \pm \sqrt{\cal D}\right)\right)~. \nonumber
\end{eqnarray}
Here ${\cal D}= \gamma^{2}\left(\nu+1/2+\gamma^{2}/4\right) +\theta^{2}(x)(1/2-\lambda)^{2}$ and the index $L$ ($R$) refers to the $x<0$ ($x>0$) values of respective quantities. To satisfy the boundary conditions at the magnetic interface and to obtain the spectrum of spin snake states, we form a linear combination of two independent bulk solutions (\ref{bulk}) as $\psi(x)=a_{L,R} \chi_{+}^{L,R}(\xi) + b_{L,R} \chi_{-}^{L,R}(\xi)$ and obtain the coefficients $a_{L,R}$ and $b_{L,R}$ from the continuity of the new spinor wave function $\psi(x)$ and its derivative $\psi^{\prime}(x)$ at $x=0$. This $4\times4$ eigenvalue problem has a nontrivial solution if the respective determinant vanishes, which leads to the following {\it exact dispersion equation} for the spin snake states
\begin{widetext}
    \begin{eqnarray} \label{det}
     \text{det}\left\vert 
     \begin{array}{c c c c}
           D_{\mu^{R}_{+}}(\xi^{R})  &  D_{\mu^{R}_{-}}(\xi^{R})  &  -c^{L}_{+} D_{\mu^{L}_{+}-1}(-\xi^{L})  & 
           -c^{L}_{-} D_{\mu^{L}_{-}-1}(-\xi^{L}) \\
           c^{R}_{+} D_{\mu^{R}_{+}-1}(\xi^{R}) & c^{R}_{-} D_{\mu^{R}_{-}-1}(\xi^{R})  & -D_{\mu^{L}_{+}}(-\xi^{L}) & 
           -D_{\mu^{L}_{-}}(-\xi^{L})\\
           D^{\prime}_{\mu^{R}_{+}}(\xi^{R})  &  D^{\prime}_{\mu^{R}_{-}}(\xi^{R}) &  
           -c^{L}_{+} D^{\prime}_{\mu^{L}_{+}-1}(-\xi^{L})  &  -c^{L}_{-} D^{\prime}_{\mu^{L}_{-}-1}(-\xi^{L})\\
           c^{R}_{+} D^{\prime}_{\mu^{R}_{+}-1}(\xi^{R}) & c^{R}_{-} D^{\prime}_{\mu^{R}_{-}-1}(\xi^{R})  &
           -D^{\prime}_{\mu^{L}_{+}}(-\xi^{L}) & -D^{\prime}_{\mu^{L}_{-}}(-\xi^{L}) \\
     \end{array}\right\vert_{x=0}=0~.
  \end{eqnarray}
\end{widetext}
The energy dispersions $\eps_{lpq}$ are described by the main Landau quantum number $l=0,1,2,\dots$ and {\it the spin-parity}, $p=\pm1$, and {\it spin-chirality}, $q=\pm1$, quantum numbers. In an antisymmetric magnetic field ($B_u=0$) the electron system coupled with SOI, possesses a spin-parity symmetry and the eigenvalue problem reduces to two $2\times2$ matrix equations. Then, for a given $l$ and $q$  the spectrum becomes degenerate in the bulk with respect to the spin-parity $p$ and exhibits a single crossing point near the magnetic interface. Introducing an additional homogeneous field brakes down the spin-parity, resulting in a lifting of the degeneracy in the bulk and in a {\it crossing-anticrossing} conversion near the magnetic interface.

Using Eq.~(\ref{det}) we carry out actual calculations of the spectrum of spin snake states in 2DES residing in InAs quantum wells with $\hbar\alpha=112.49$ meV {\AA}, $g=-14.8$, and $m^\ast=0.026 m_e$ where $m_e$ is the free electron mass. Notice that both the Rashba and Zeeman fields can be significant in these structures, however, in relatively weak magnetic fields the SOI dominates, {\it i.e.} $\gamma\gg \lambda$. 
\begin{figure}[t]
  \includegraphics[height=6cm]{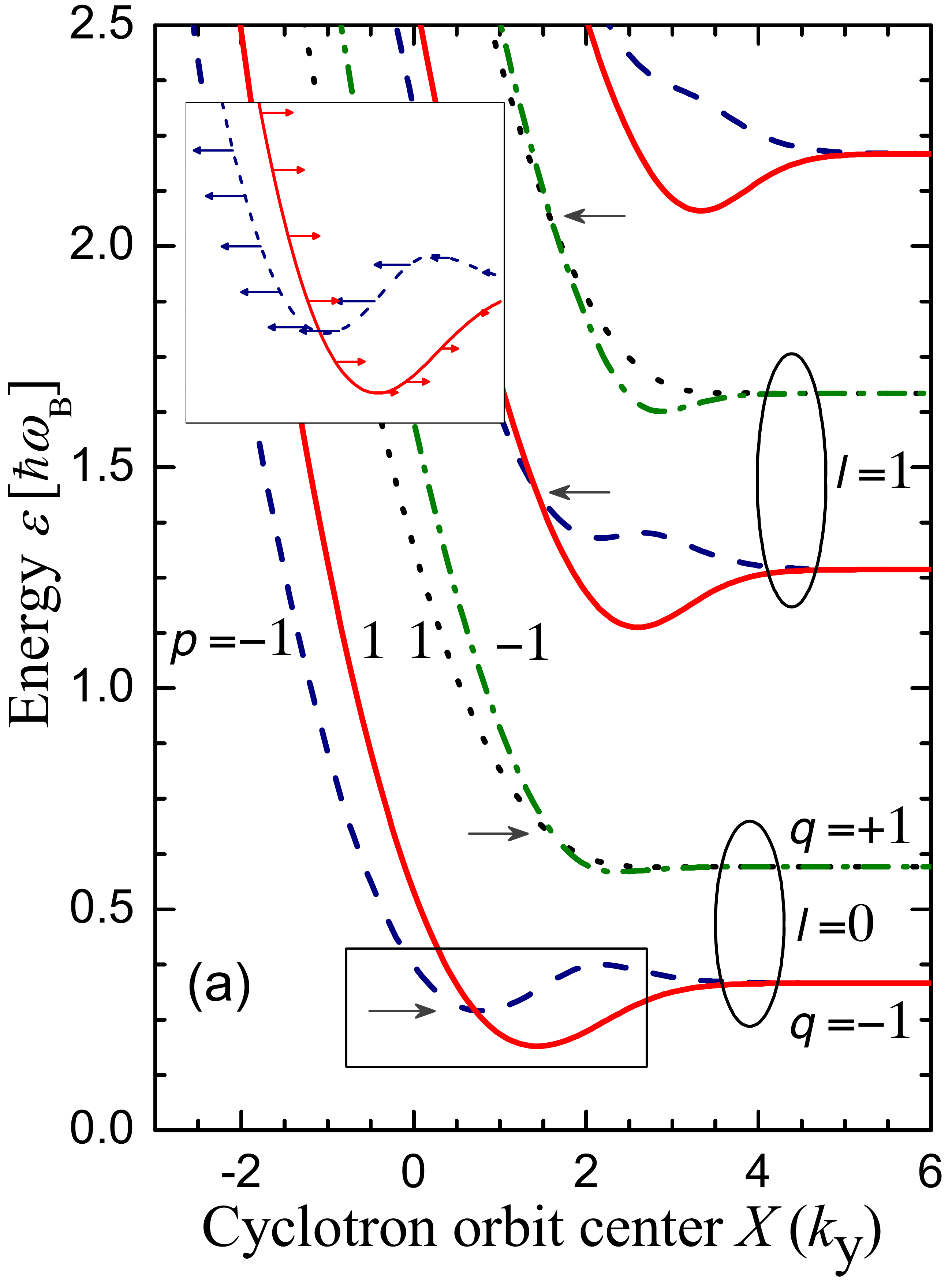}
  \includegraphics[height=6cm]{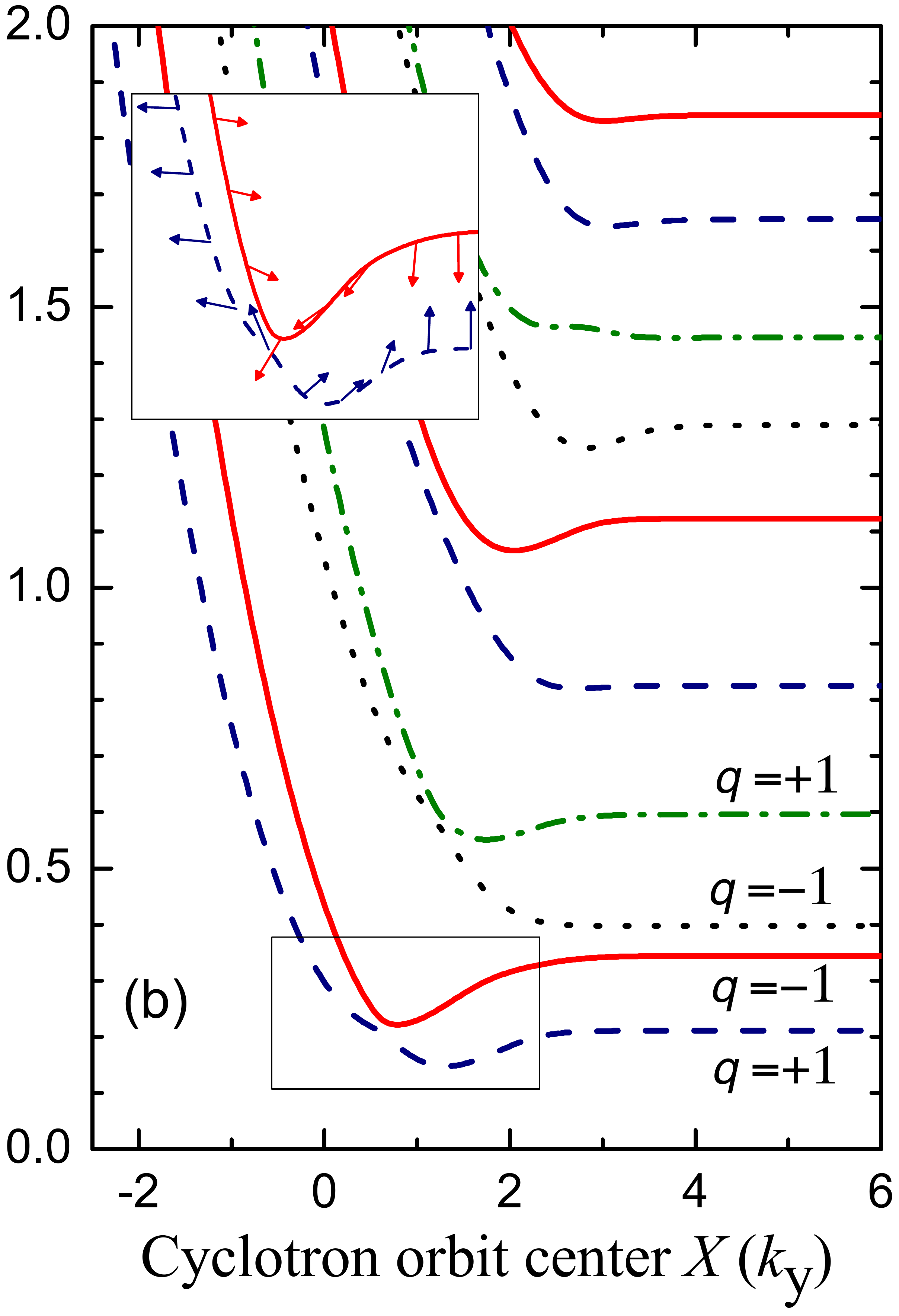}
\caption{Energy dispersions of spin snake states as a function of the position of the cyclotron orbit center for: (a) $B_u=0$ and (b) $B_u=0.2B_0$. For given Landau number $l=0,1,2\dots$ the four energy branches are described by the spin-parity, $p=\pm1$, and the spin-chirality, $q=\pm1$, quantum numbers and are depicted, respectively, by solid, dashed, dotted, and dash-dotted curves. Insets are zooms of the open rectangles. In the main figure the arrows indicate the crossing points of the even and odd spin-parity energy terms while in the insets the arrows depict the electron average spin. The spin-orbit and Zeeman field parameters are $\gamma \approx 0.3$ (for $B_{u}=0$) and $\lambda \approx 0.1$.}
\label{disp}
\end{figure}

Fig.~\ref{disp} shows the energy spectrum of snake states, $\eps_{lpq}(k_{y})$, as a function of the cyclotron orbit center position $X(k_{y})$. Introducing the spin-orbit and the Zeeman fields in the presence of a magnetic interface removes the degeneracy of the usual Landau structure in a specific way. Groups of four different energy levels are formed as shown in the figure by the different dispersion curves, for different main Landau number $l$. In an antisymmetric magnetic field ({\it cf.} Fig.~\ref{disp}(a)) the energy branches far from the magnetic interface represent the spin-split quasi-bulk Landau levels, which are described by the $q=\pm 1$ spin-chirality for given $l$. The splitting energy is determined by the strength of the spin-orbit and Zeeman fields. Each quasi-bulk Landau level remains partially degenerate due to the spin-parity symmetry. Usually, a magnetic field breaks the spin-parity \cite{Debald2005}, here, however, the combined action of the magnetic interface and the SOI preserves the spin-parity in the bulk. Close to the magnetic interface, snake orbits with even ($p=1$) and odd ($p=-1$) spin-parities start to hybridize, which removes the degeneracy of the spin snake states except in one singular point where the even and odd terms cross. It is seen that the application of an external uniform magnetic field ({\it cf.} Fig.~\ref{disp}(b)) brakes the spin-parity symmetry$-$the singular crossing points are converted into anticrossings and the branches of the spectrum become fully spin-resolved. In this case, the energy of the quasi-bulk Landau levels is additionally determined by the two different cyclotron frequencies and this is an extra source of asymmetry in the spectrum. Notice that for given $l$ the electron Larmor radius of classical orbits depends on the  $p$ and $q$ quantum numbers of the snake states. 


\begin{figure}[t]
   \includegraphics[width=\columnwidth]{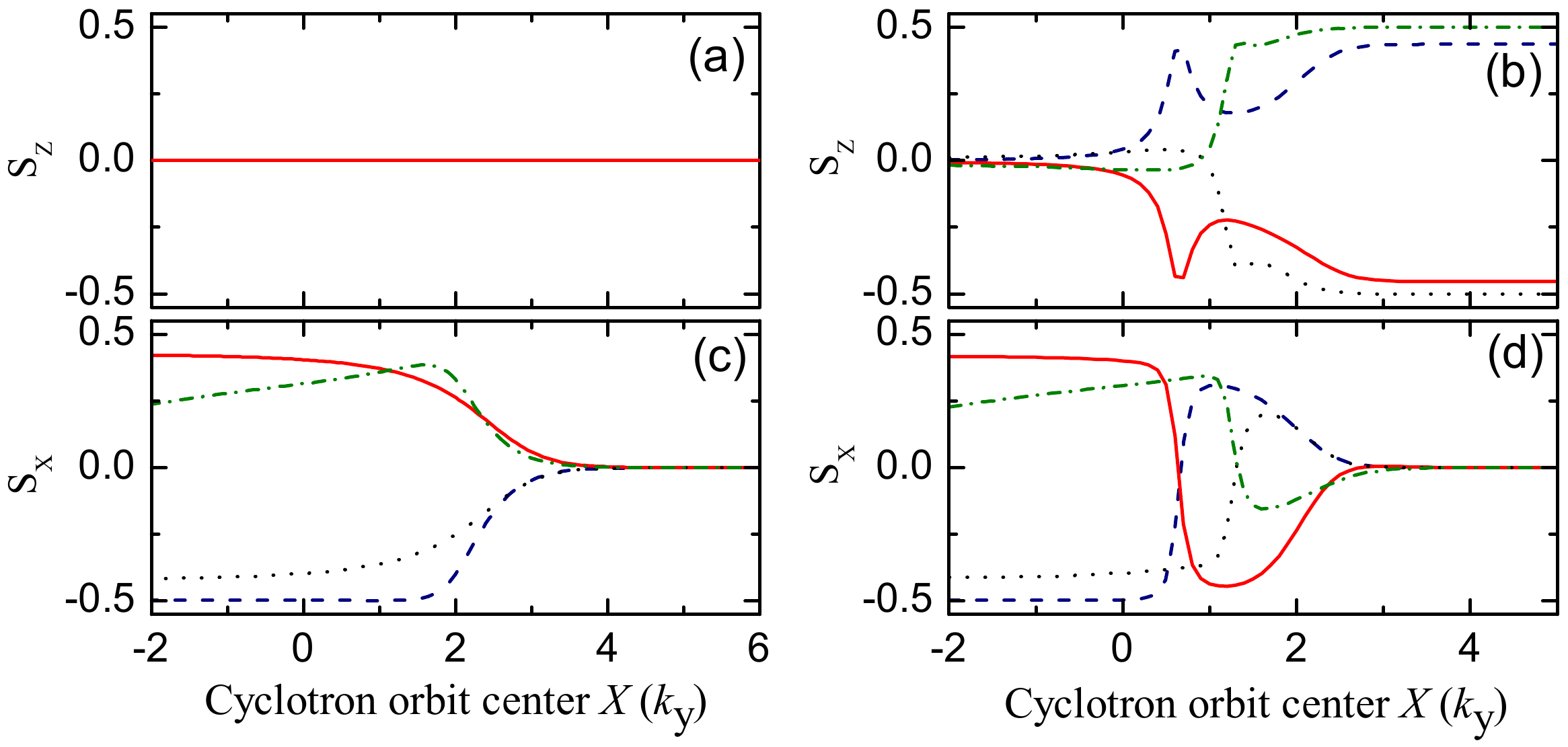}
   \caption{The $S_{z}$ and $S_{x}$ spin projections of the snake states for the lowest $l=0$ Landau band as a function of the position of the cyclotron orbit center. (a) and (c) are for $B_u=0$ and (b) and (d) are for $B_u=0.2B_0$. The other parameters and the type of curves corresponding to the $p=\pm1$ and $q=\pm1$ states are the same as in Fig.~\ref{disp}}.  
\label{spin}
\end{figure}

\begin{figure*}[t]
  \includegraphics[width=.9\columnwidth]{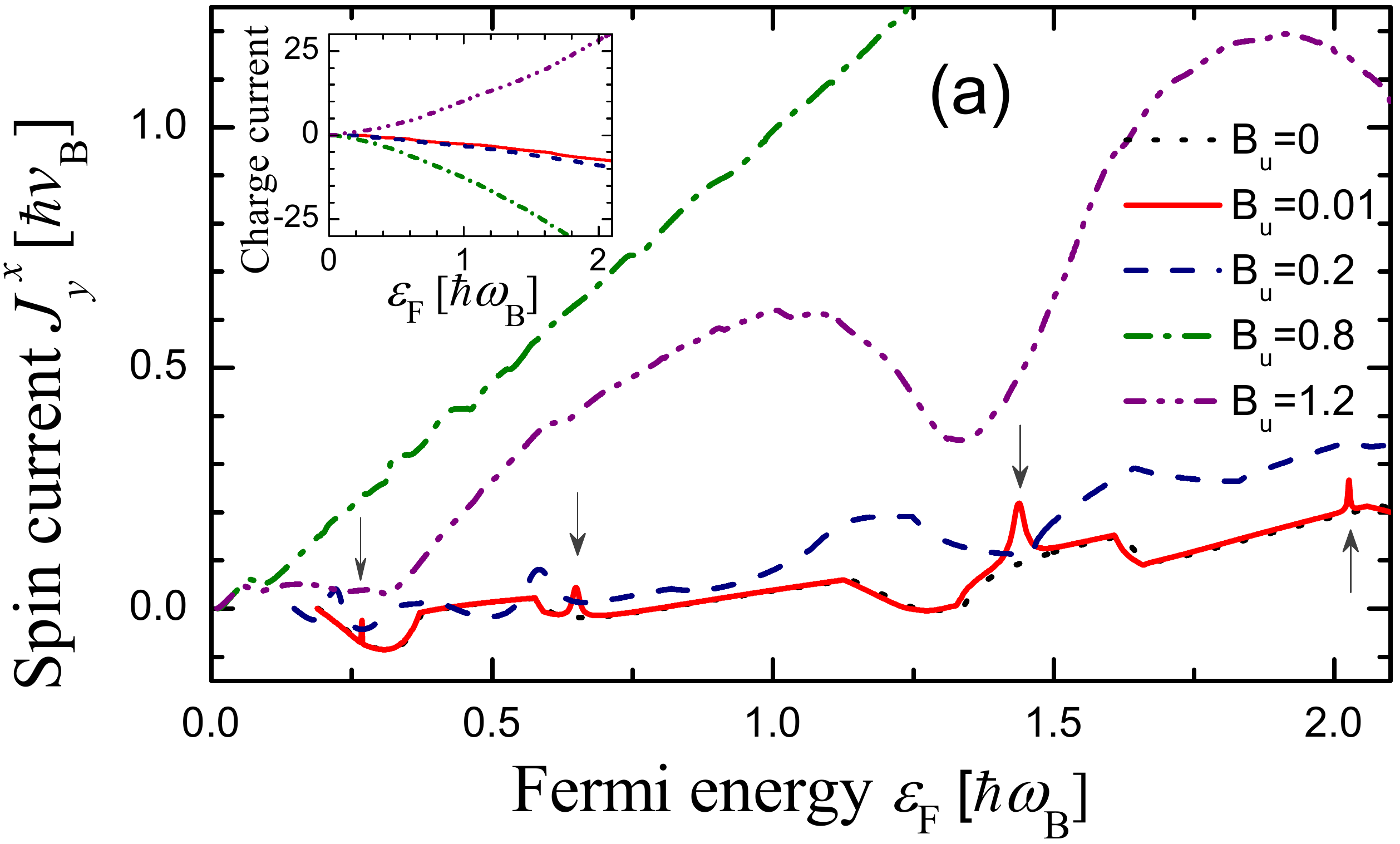} \hspace{5mm}
  \includegraphics[width=.9\columnwidth]{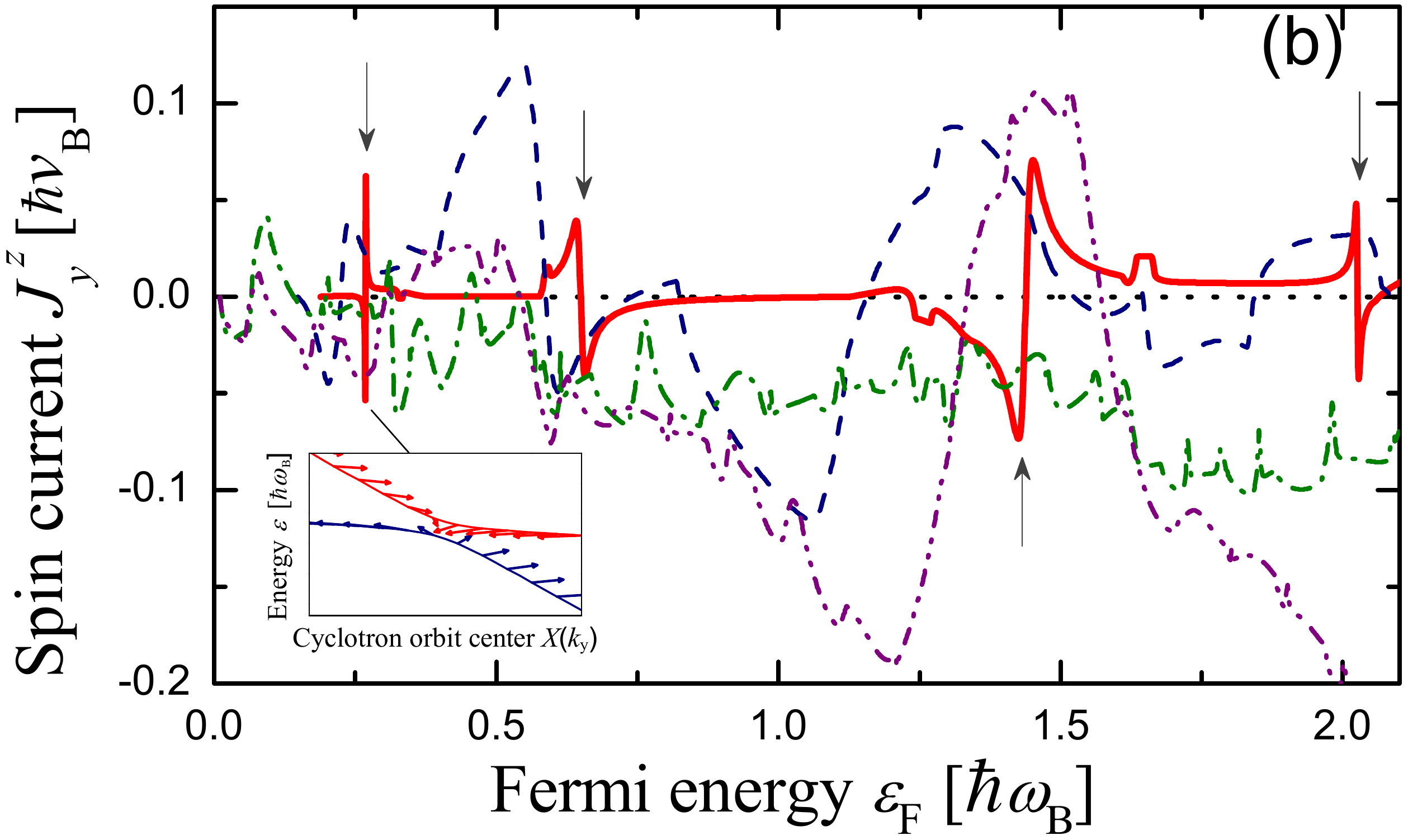}
  \caption{The $J^{x}_{y}$ and $J^{z}_{y}$ currents of the respective spin projections along the magnetic interface, shown respectively in (a) and (b), as a function of the Fermi energy. The dotted, solid, dashed, dot-dashed, and dot-dot-dashed curves are calculated, respectively, for $B_{u}=0, 0.01, 0.2, 0.8$ and $1.2B_{0}$. The other parameters are the same as in Fig.~\ref{disp}. The arrows show the positions of the local beatings and peaks, corresponding to the crossings in the spectrum of spin snake states in Fig.~\ref{disp}(a). Inset in (a) plots the corresponding charge currents. Inset in (b) shows an anticrossing behavior induced in the spectrum of spin snake states by an external magnetic field with arrows indicating the spins. }
\label{SPC}
\end{figure*}

As seen in Fig.~\ref{spin} the combined action of the magnetic interface and the spin interactions has critical consequences for the behavior of the electron average spin projections, $S_{x,z}(k_{y})=\int dx \psi^{+}(x)\hat\sigma_{x,z}\psi(x)$, which are calculated using the energy dispersions $\eps_{lpq}(k_{y})$. We observe that in an antisymmetric magnetic field ($B_{u}=0$) the average spin vanishes in the bulk for any values of the spin-parity and the spin-chirality quantum numbers. The $S_x$ projection of the spin vanishes in the limit of $X(k_{y})\rightarrow +\infty$ because of the electron circular motion perpendicular to the magnetic field. We find that the $S_z$ component of the average spin also vanishes independently of the electron position with respect to the magnetic interface ({\it cf.} Fig.~\ref{spin}(a)). This implies that an electron becomes a nonlocal composite particle, which captures simultaneously the spatially separated negative and positive magnetic flux quanta and manifests itself as a particle with zero spin. The nonlocality originates from the cyclotron rotations in the opposite directions of the composite particle on both sides of the magnetic interface, respectively, with the up and down entangled projections of the spin. 
The application of an external homogeneous magnetic field destroys this spin-parity symmetry and the $S_{z}$ projection of the spin jumps from zero to its quantized value. This jump is strongly pronounced at the crossing points, $X_{c}$, of the energy branches with the even and odd spin-parity  ({\it cf.} the peaked behavior of $S_{z}$ at $X_{c}\approx 0.71$ for the solid and dashed curves in Fig.~\ref{spin}(b)). 
In the opposite limit of $X(k_{y})\rightarrow -\infty$ the snake states exhibit an unilateral motion along the magnetic interface and $S_{z}$ vanishes. As seen in Fig.~\ref{spin}(c), being zero in the bulk the $S_{x}$ spin projection becomes finite when the electron moves towards the interface.  This behavior is modified in a finite external homogeneous field ({\it cf.} Fig.~\ref{spin}(d)). It is seen that close to the interface $S_{x}$ starts to oscillate before reaching its quantized value. Notice also that the orientation of $S_{x}$ is independent of the  direction of the total averaged magnetic field, which determines, however, the orientation of the $S_{z}$ projection.

An external homogeneous magnetic field brakes the spin-parity symmetry, removing not only the degeneracy of the quasi-bulk $p=\pm1$ Landau levels with a vanishing velocity but, what is more important for transport, gapping out spectral branches of the extended snake states at the crossing points  $X_{c}$ close to the magnetic interface ({\it cf.} the inset of Fig.~\ref{SPC}(b)). Thus, the combined effect of the magnetic interface and the spin interactions induces a helical liquid of spin snake orbits with additional channels for carrying an uncompensated spin current. The overall spin current is calculated from $J^{x,z}_y=\sum \left. S_{x,z}(k_y) v_y(k_y) \right|_{\eps_{lpq}(k_{y})=\eps_{F}}$ where the sum is over the quantum numbers of current carrying spin snake channels at the Fermi energy, $\eps_{F}$, and the electron group velocity along the magnetic interface is $v_y(k_y)=\partial\eps_{lpq}(k_{y})/\partial k_{y}$. In Fig.~\ref{SPC} we observe that the application of an external uniform magnetic field induces local peaks and beatings, respectively, in the current of the $S_{x}$ and $S_{z}$ spin projections versus the Fermi energy (compare the curves corresponding to $B_u=0$ and $B_u=0.01 B_0$), which appear precisely at the crossing points of the energy levels with even and odd spin-parity. Notice that the charge current ({\it cf.} inset in Fig.~\ref{SPC}(a)) does not exhibit such beatings or peaks instead a small oscillatory structure is not visible on the scale of its variation. When increasing the applied uniform field the shape of the energy dispersion changes and the beatings and peaks are broadened. As seen in Fig.~\ref{SPC}(a), for $0<B_{u}<B_{0}$ the overall magnitude of the $J^{x}_{y}$ spin current increases monotonically with $B_{u}$. For $B_{u}=0.8B_{0}$ it shows almost a linear dependence on the Fermi energy because the velocity of extended snake orbits is linear in $\eps_{F}$ while the x-component of the average spin, $S_{x}$, is approximately directed along the $x$ axis. However, this monotonic tendency in the dependence of $J^{x}_{y}$ versus $B_{u}$ ceases to exist at $B_{u}=B_{0}$. For $B_u>B_0$ the magnetic field does not change its sign at the magnetic interface, the extended snake orbits are transformed into the spiral orbit states with a smaller velocity and this results in a suppression of the spin current, which is observed in Fig.~\ref{SPC}(a) for $B_{u}=1.2 B_{0}$.

In conclusion, we demonstrated that in an antisymmetric magnetic field carriers in the spin snake states are rendered into nonlocal composite particles, originating from the spatially separated entangled spins. Applying an external uniform magnetic field destroys the spin-parity symmetry gapping out the spectral branches of the spin snake states with finite velocities. This leads to new structures in the behavior of the spin current. The predicted effects are of experimental relevance for exploring the helical liquid of snake states in quantum entanglement and spin transport phenomena to generate novel information processing and spintronic applications.

This work was supported by the Methusalem program of the Flemish government and the Flemish Science Foundation (FWO-Vl).

\end{document}